\documentclass[twocolumn]{el-author}

\newcommand{\ua}{\uparrow}
\newcommand{\nc}{\newcommand}
\nc{\da}{\downarrow} \nc{\hc}{\hat{c}} \nc{\hS}{\hat{S}}
\nc{\bra}{\langle} \nc{\ket}{\rangle} \nc{\eq}{equation (\ref}
\nc{\h}{\hat} \nc{\hT}{\h{T}}\nc{\be}{\begin{eqnarray}}
\nc{\ee}{\end{eqnarray}}\nc{\rd}{\textrm{d}}\nc{\e}{eqnarray}\nc{\hR}{\hat{R}}\nc{\Tr}{\mathrm{Tr}}
\nc{\tS}{\tilde{S}}\nc{\tr}{\mathrm{tr}}\nc{\8}{\infty}\nc{\lgs}{\bra\ua,\phi|}\nc{\rgs}{|\ua,\phi\ket}
\nc{\hU}{\hat{U}}\nc{\lfs}{\bra\phi|}\nc{\rfs}{|\phi\ket}\nc{\hZ}{\hat{Z}}\nc{\hd}{\hat{d}}\nc{\mD}{\mathcal{D}}
\nc{\bd}{\bar{d}}\nc{\bc}{\bar{c}}\nc{\mc}{\mathcal}\nc{\ea}{eqnarray}\nc{\mG}{\mathcal{G}}\nc{\bce}{\begin{center}}
\nc{\ece}{\end{center}}
\date{XX XX 2019}

\begin{document}

\title{Cryogenic low power CMOS analog buffer at 4.2K }

\author{Yajie Huang, Chao Luo, Tengteng Lu, Zhen Li, Jun Xu and Guoping Guo}
	
\abstract{A novel power-efficient analog  buffer at  liquid helium temperature is proposed. The proposed circuit is based on an input stage consisting of two complementary differential pairs to achieve rail-to-rail level tracking. Results of simulation based on SMIC 0.18$\mu$m CMOS technology show the high driving capability and low quiescent power consumption at cryogenic temperature. Operating at single 1.4 V supply, the circuit could achieve a slew-rate of +51 V/$\mu$s and -93 V/$\mu$s for 10 pF capacitive load. The static power of the circuit is only 79$\mu$W.}

\maketitle

\section{Introduction}

According to past researches, the classical-type electronic controller of quantum computer is implemented at room-temperature (RT), which needs wiring and electronic interface to connect to the cryogenic quantum processor \cite{1}. However, this method brings high cost and unreliability to the control of quantum processor. Hence, a CMOS control circuit operating at cryogenic temperature has been proposed. To realize the proposed alternative solution, we have finished the characterization and modelling of SMIC 0.18$\mu$m CMOS technology at cryogenic temperature in past research \cite{2}.        
 
CMOS analog buffer is an important basic building block in many applications at room temperature. It can be used as a voltage buffer between different unit circuits or to detect the voltage value of an internal node without affecting the operation of the internal circuit. Similarly, the quantum controller circuits need voltage buffer to drive large capacitive load and track the voltage level. Moreover, this buffer needs to meet some particular criteria like high slew-rate, operating in liquid helium tank and consuming extremely low power. Class-AB buffer can solve this limitation because it features a balance between slew-rate and power consumption \cite{3}. Several class-AB buffers have been proposed. Carrillo et al. introduced a rail-to-rail CMOS analog buffer circuit which based on a differential class AB input stage \cite{4}. This design could realize rail-to-rail voltage level tracking but the increase of transistors resulted in higher power consumption. Based on Carillo's circuit, Sawigun et al. used an adaptive biasing technique to reduce the number of transistors and power consumption relative to the original circuit \cite{5}. The static power of Sawigun's circuit is 282$\mu$W which needs further power reduction for the application in 4.2K. 

In this Letter, a novel cryogenic power efficient CMOS buffer which is based on differential input pairs and class AB output stage is presented. It features low operation temperature, low static power consumption and high slew-rate. The cryogenic model of SMIC 0.18$\mu$m CMOS technology are implemented to complete the simulation of this circuit.               

\section{Cryogenic model}

At cryogenic environment, the MOSFETs will reveal new characteristics because of the freeze-out effect and kink effect. The carrier freezes out in the MOSFET channel region at cryogenic temperatures, so a higher gate drive voltage is required to inject carriers into the channel region. In addition, Table 1 illustrates the threshold voltage ($V_{th}$) change of different width-to-length ratio MOSFETs at LHT and RT. These parameters were extracted from the measurement results of various transistors and temperatures. At cryogenic temperature, threshold voltage has an significant rise and the $V_{th}$ of PMOS has a greater increase than NMOS's. Besides, Ion to Ioff ratio (turn-on and turn-off current ratio) increases at low temperatures which means the tested CMOS transistors can operate well as a switch for digital circuits at cryogenic temperatures and have low quiescent power consumption at the same time \cite{6}. Gate transconductance ($G_{m}$) represents the gate to source current control capability. With the decrease of temperature, the $G_{m}$ of MOSFET increases several times. This increase supplies a wider bandwidth for the same power budget. Measurement results of the thin-oxide SMIC 0.18$\mu$m CMOS technology transistors had different performance with the RT BSIM3V3 model \cite{2,7}. To eliminate the deviation, a compact model based on BSIM3V3 has been proposed by using default parameters. In this Letter, the results and SPICE model are used to the design and simulation of the buffer circuit to obtain a better performance at cryogenic temperature. 

\begin{table}[h]
\processtable{The threshold voltage of different width-to-length ratio 1.8V N/PMOS at LHT and RT}
{\begin{tabular}{|l|l|l|}\hline
Temperature & 4.2K & 298K\\\hline
NMOS 10/10 ($\mu$m) &  536mV &  448mV\\\hline
NMOS 10/0.6 ($\mu$m) &  608mV &  501mV\\\hline
NMOS 0.22/0.18 ($\mu$m) &  657mV &  516mV\\\hline
PMOS 10/10 ($\mu$m) & -801mV & -461mV\\\hline
PMOS 10/0.6 ($\mu$m) & -792mV & -481mV\\\hline
PMOS 0.22/0.18 ($\mu$m) & -832mV & -507mV\\\hline
\end{tabular}}{}
\end{table}
 
\section{Proposed circuit}

Fig. 1 shows the transistor level implementation of the proposed cryogenic CMOS buffer. The circuit is made up of two complementary differential pairs and a output stage. The differential pairs use current-mirrors connection to increase the current of output node and make the output voltage doubled. As a consequence, the gain of this pairs could be same as the gain of dual output nodes differential amplifier. This circuit utilizes self-biasing scheme(the gate of M1 connected to the gate of M4 and the gate of M10 connected to the gate of M6) rather than external reference source to generate bias current.

\begin{figure}[h]
\centering{\includegraphics[width=90mm]{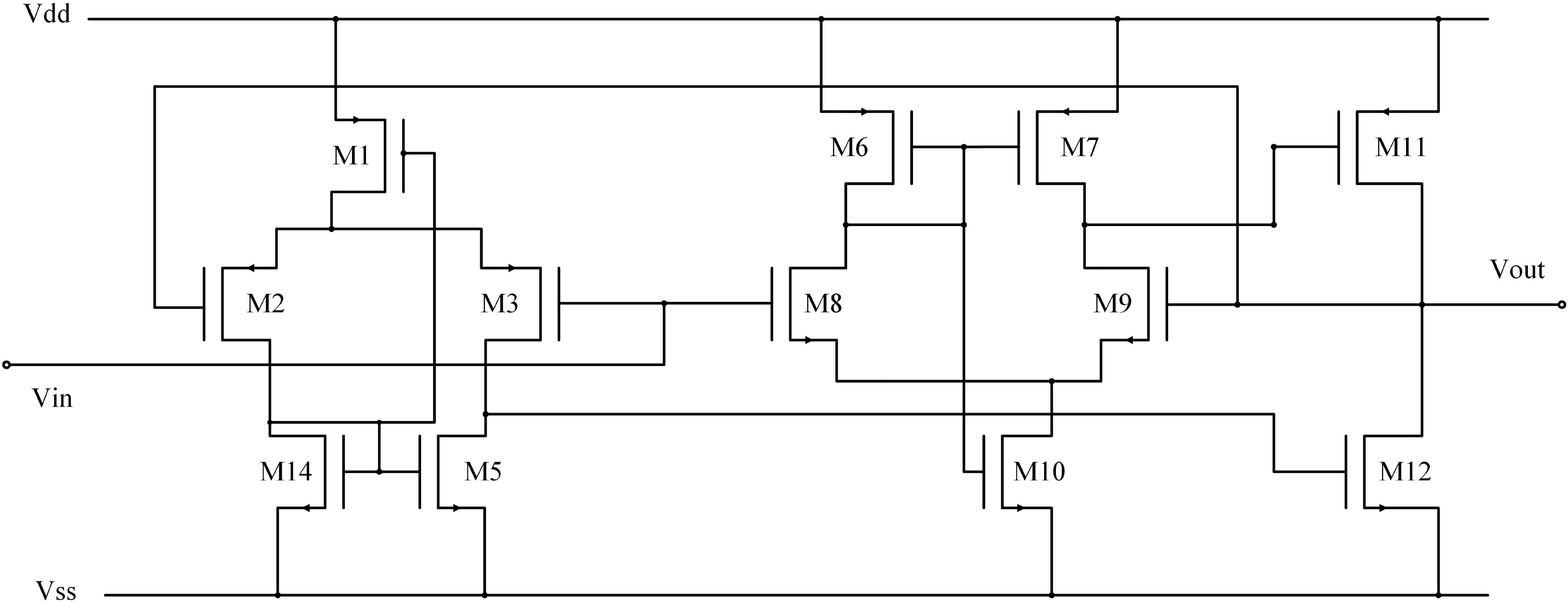}}
\caption{Circuit of the proposed cryogenic buffer
\source{Transistor sizes W/L (in $\mu$m/$\mu$m ) were 
	PMOS: 4/0.2 (M1,M2,M3,M11), 6/0.2 (M6,M7)}
\source{NMOS:2/0.2 (M4,M5,M8,M9,M10,M12)}}\label{kondodotresistance}
\end{figure}

The operation of this buffer can be divided into three parts.  When the input signal $V_{in}$ approaches $V_{ss}$, the NMOS pair cuts off and the current mirror M6-M7 does not send any current contribution to the output stage. However, M4-M5 turns on, transporting a current from M2 to output branch to maintain the buffer turned on. Similarly, when $V_{in}$ is close to $V_{dd}$, the PMOS pair cuts off and the NMOS pair starts working to keep the circuit turned on. If the input signal $V_{in}$ is in the mid-supply region, both pairs are active and send current to the output node. Thus, whether the input signal is tiny or close to $V_{dd}$, the buffer can always provide equivalent gain. At the same time, the high drive capacity of the class AB output stage would improve the dynamic performance of the proposed circuit.   
 
Power consumption is a principal indicator of low temperature circuit design. By cryogenic parameters extraction and modelling , the transistor parameters affecting the analogue circuit performance strongly all get enhanced. These improvements allow cryogenic circuits to achieve semblable performance but have lower power consumption comparing to room temperature circuits. Take the NMOS pair as an example. Assume that the large signal input is zero and only tiny signal connects to the input port. Due to the small signal $v_{gs8}$ added to M8, the DC current of M8 is $I_{bias}/2$. Thus the total current of M8 could be expressed as:
\begin{align}
	i_{D8}=I_{D8}+i_{d8}=i_{o}+g_{m8}v_{gs8}
\end{align} 
Likewise, the total current of M9 is:
\begin{align}
	i_{D9}=I_{D9}+i_{d9}=\frac{I_{bias}}{2}+g_{m9}v_{gs9}
\end{align}
Since the current mirror forces the current to flow equal, the small signal output current equals the difference between $i_{D7}(=i_{D8})$ and $i_{D9}$:
\begin{align}
	i_{o}=g_{m9}v_{gs9}-g_{m8}v_{gs8}
\end{align}
The NMOS differential pair output voltage is: 
\begin{align}
	v_{o}=v_{gs8}-v_{gs9}
\end{align}
M8 and M9 are completely symmetric so the $i_{o}$ is given by following:
\begin{align}
	i_{o}=-g_{m8}v_{o}
\end{align} 
Therefore the differential pair transconductance $G_{md}=\frac{I_{o}}{v_{o}}=-g_{m8}$. By this token, the transconductance of differential pair is determined by the transconductance of the input transistor. At cryogenic temperature, the MOSFETs' $G_{m}$ increases obviously which means that for the same power budget the circuits have a wider bandwidth at liquid helium temperature \cite{2}. As for buffer circuit, operating at cryogenic temperature could get analogous performance but consume less power.

\section{Simulation results}

\begin{figure}
	\centering{\includegraphics[width=70mm]{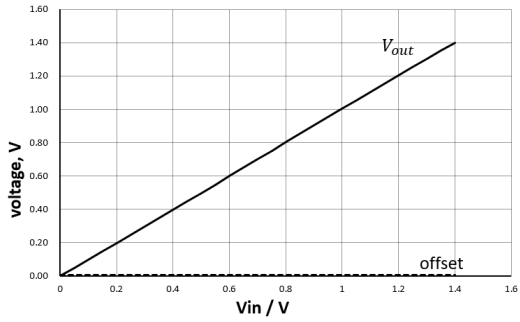}}
	\caption{DC transfer characteristic of cryogenic buffer in Fig. 1
		\source{solid line output voltage}
		\source{dash line offset voltage}}\label{kondodotresistance}
\end{figure}

\begin{figure}
	\centering{\includegraphics[width=70mm]{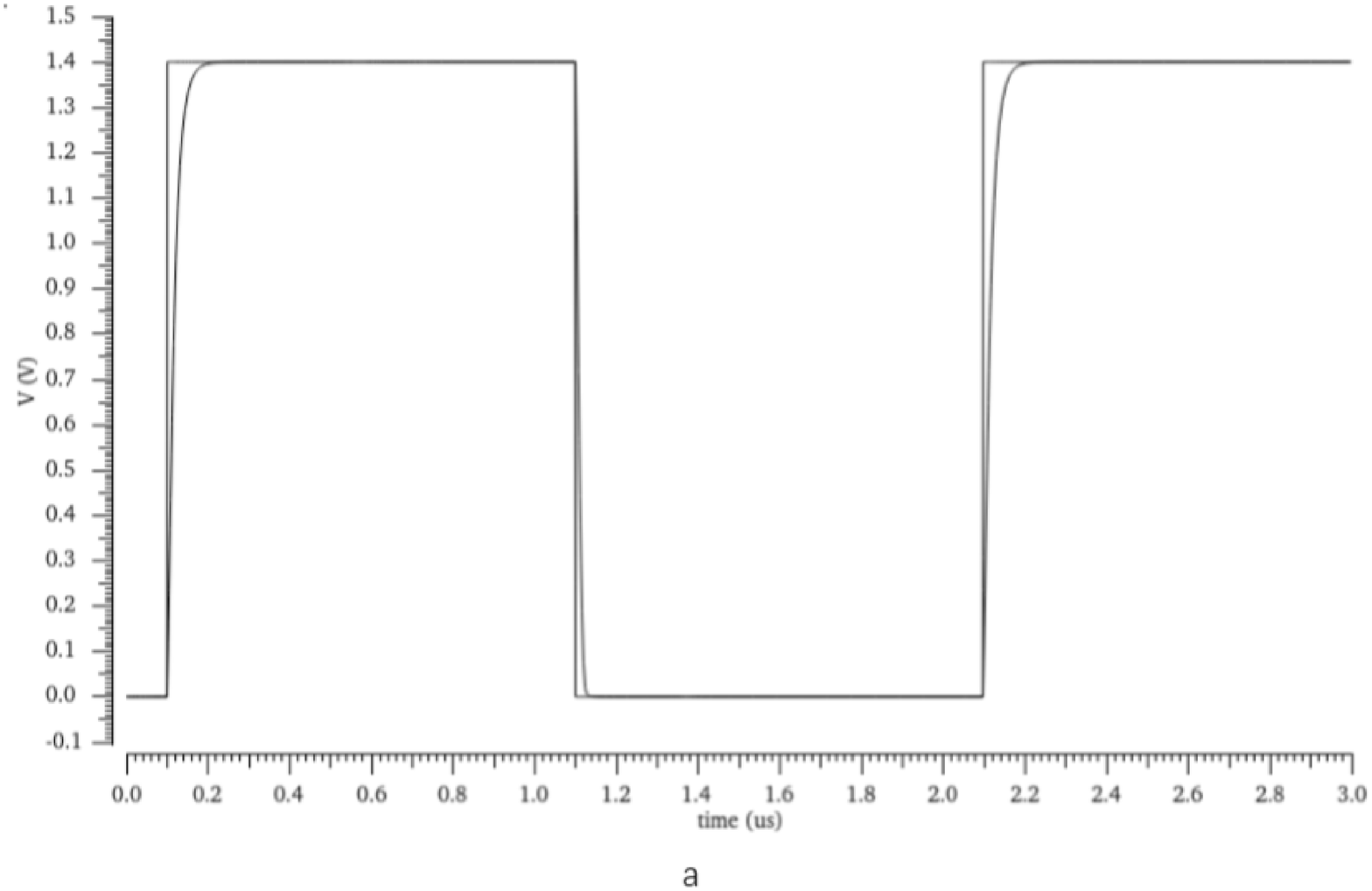}
		\includegraphics[width=70mm]{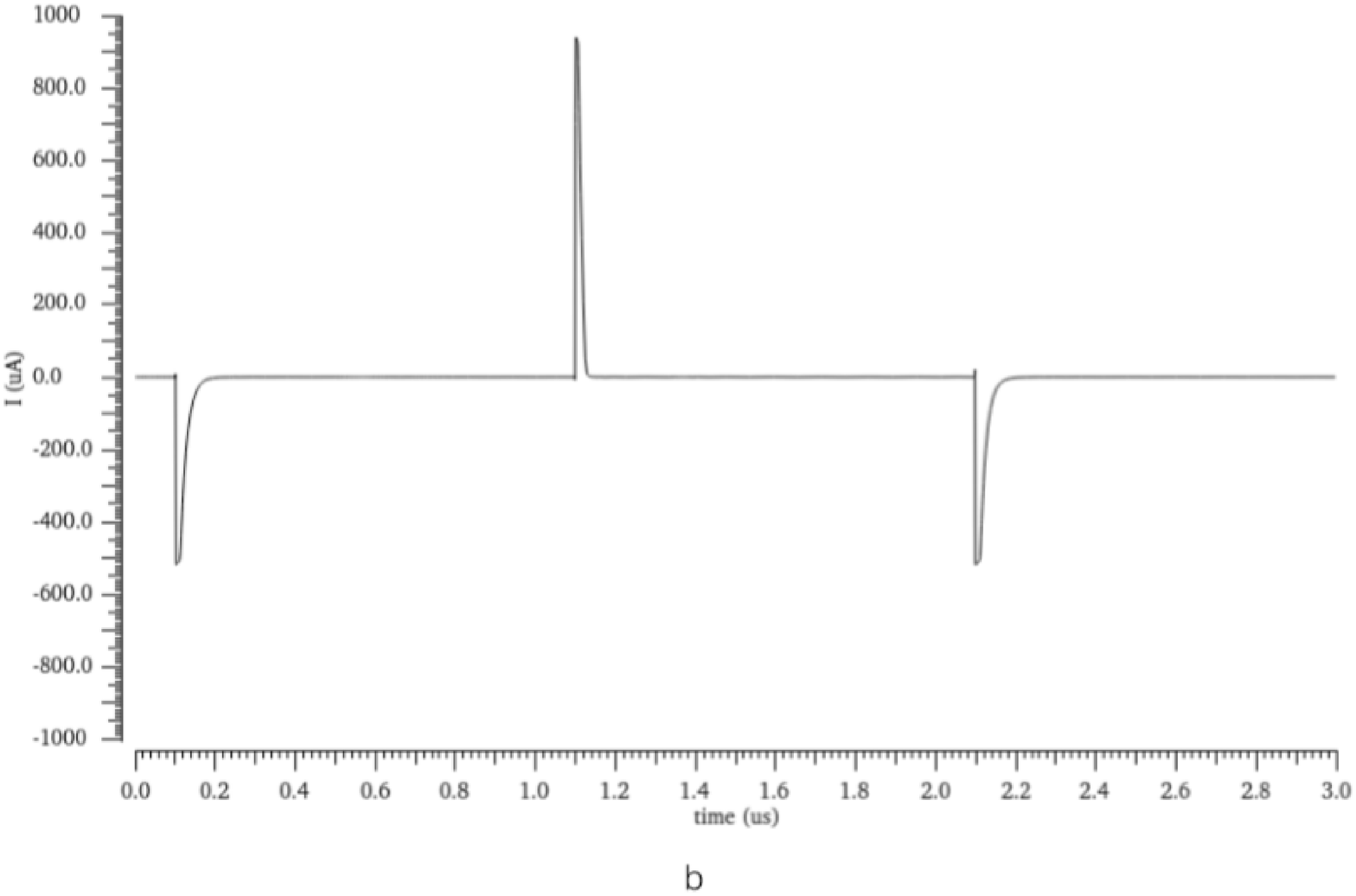}}
	\caption{Transient response of cryogenic buffer for 1.4V 500 kHz square input signal with 10 pF load   
		\source{\textit{a}  Input and output voltages}
		\source{\textit{b}  Current through output transistor}}\label{kondodotresistance}
\end{figure}

The proposed analog buffer in Fig. 1 was designed in SMIC 0.18$\mu$m CMOS technology and the SPICE model was modified to be consistent with the cryogenic performance of MOSFETs. The circuit operated with a single 1.4V supply and a capacitive load of 10pF.

Fig. 2 shows the DC characteristic of the cryogenic buffer together with the offset voltage. As illustrated, the buffer circuit possesses rail-to-rail drive capability. Fig. 3 shows the large signal transient response of the circuit to 1.4V 500 kHz input square signal with 10 pF load capacitance. As expected, the circuit achieves a high slew-rate with low power consumption due to the class AB operation of the output stage.

At last, Table 2 summarizes some typical parameters related to the performance at different temperature of the proposed buffer. Taking low temperature parameter as an example, the simulated open-loop gain and unity-gain frequency are 59 dB and 199 MHz. The simulation results show a total harmonic distortion (THD) equal to -52.7 dB for a 1.4 $V_{pp}$ 100 kHz input sinewave signal. The positive and negative slew-rate of this buffer are +51 V/$\mu$s and -93V/$\mu$s. Apparently, this buffer achieves approximate performance with less quiescent power consumption at liquid helium temperature compared to room temperature. The differences of unity-gain frequency and slew-rate are mainly resulted from the increase of $G_{m}$ and the decrease of the operation current of MOSFETs at cryogenic temperature.

\begin{table}[h]
\processtable{Simulated performance of the analogue buffer
		($V_{DD}$=1.4V, $C_{L}$=10pF, T=4.2K, 298K)}
{\begin{tabular}{|l|l|l|}\hline
Parameter & 4.2K & 298K\\\hline
Open-loop gain &  59 dB & 59 dB\\\hline
Unity-gain frequency &  199 MHz & 137 MHz\\\hline
Static power consumption &  79 $\mu$W & 138 $\mu$W\\\hline
THD(1.4$V_{p}$@100 kHz) & -52.7dB & -47.6dB\\\hline
$SR^{+}$ / $SR^{-}$ (V/$\mu$s) & 51 / 93 & 54 / 71\\\hline
\end{tabular}}{}
\end{table}

\section{Conclusion}

A rail-to-rail voltage buffer operating at 4.2K is proposed. The circuit adopts two complementary differential input pairs. As a result, the buffer can provide equally gain as the input voltage approaches the supply voltage. Meanwhile, the class AB output stage leads the circuit to a low power consumption and a high slew-rate. Consequently, the buffer could be used to drive large capacitive loads. Simulation results based on LHT SPICE model indicate that this buffer can operate at liquid helium temperature and possess less power budget without diminishing its drive capability.   
\vskip3pt
\ack{The authors would like to thank SMIC for device
	fabrication and software support. This work was supported by the
	National Basic Research and Development Program of China, China
	(Grant No.2016YFA0301700), the National Natural Science Foundation
	of China, China (Grant No.11625419), the Anhui initiative in Quantum
	information Technologies, China (Grants No.AHY080000). This work
	was partially carried out at the USTC Centre for Micro- and Nanoscale
	Research and Fabrication.}

\vskip5pt

\noindent Yajie Huang, Chao Luo, Tengteng Lu, Zhen Li, Jun Xu and Guoping Guo (\textit{Key
	Laboratory of Quantum Information, University of Science and
	Technology of China, Hefei, Anhui 230026, People’s Republic of China})
\vskip3pt

\noindent Yajie Huang (\textit{Origin Quantum Computing Company Limited, Hefei, Anhui, 23088, China})
\vskip3pt

\noindent E-mail:  gpguo@ustc.edu.cn

\end{document}